\documentclass[%
superscriptaddress,
groupedaddress,
showpacs,
 amsmath,amssymb,
 aps,
pra,
floatfix,
]{revtex4-2}

\usepackage{graphicx,url}

\begin{document}

\title{Photoionization-induced reflection for benchmarking of the photoionization models in solids}

\author{A. Husakou}
\email{gusakov@mbi-berlin.de}
\affiliation{Max Born Institute, Max Born Str. 2a, D-12489 Berlin, Germany}

\author{Z. Ruziev}
\affiliation{Tashkent State Technical University,\\ University street 2, 100097\\
Tashkent, Uzbekistan}
\affiliation{University of Geological Sciences,\\ Olimlar street 64, 100164\\ Tashkent, Uzbekistan}

\author{K. Koraboev}
\affiliation{Tashkent State Technical University,\\ University street 2, 100097\\
Tashkent, Uzbekistan}
\affiliation{University of Geological Sciences,\\ Olimlar street 64, 100164\\ Tashkent, Uzbekistan}

\author{F. Morales}
\affiliation{Max Born Institute, Max Born Str. 2a, D-12489 Berlin, Germany}

\author{M. Richter}
\affiliation{Max Born Institute, Max Born Str. 2a, D-12489 Berlin, Germany}

\author{K. Yabana}
\affiliation{Tsukuba University, 1 Chome-1-1 Tennodai, \\Tsukuba, Ibaraki 305-8577, Japan}

\begin{abstract}
The choice of the most suitable analytic photoionization model in solids is a challenging task with no default solution. Here we show how the best formalism can be determined based on the waveform of the pulse reflected by a sample due to photoionized almost-free electrons in the conduction band. For a typical case of diamond, we compare three simple models and benchmark them against highly accurate first-principle TDDFT simulation, by analysing the fit between the reflected pulses in time and frequency domain. For the aims of this paper, we have developed a software package, called PIGLET, for the FDTD simulation of photoionization-governed propagation, which is now freely available for the scientific community. Furthermore, we show that due to interband contributions for very short sub-10-fs pulses a semi-classical description based on any analytical photoionization model fail to provide an adequate description.  
\end{abstract}

\maketitle

\section{Introduction}
In the last decades, interaction of strong light with solid-state materials accompanied by electron transition to the conduction zone has attracted a significant attention. This is motivated by recent progress in various areas such as material modification \cite{app_mod}, high harmonic generation \cite{app_hhg}, petahertz electronics \cite{app_pe1,app_pe2}, all-optical sampling \cite{app_sam}, control of currents in solid state \cite{app_ctr}, short pulse generation \cite{app_our}, creation of nanoparticles by laser ablation \cite{Sem_Ng} and so on. The accurate simulation of electron transition rates from the valence band to the conduction band(s) is of paramount importance for such applications, since it determines, among others, the nature of the light interaction with the dielectric. In particular, for a sufficient (critical) concentration of conduction-band electrons, metallization can occur, with dielectric function of the material becoming negative due to almost-free-electron plasma contribution, leading to light reflection.  

For the description of the transition of the electrons from the valence band to the conduction band in dielectrics, several accurate and more or less first-principle methods are available. They include many-body perturbation equations \cite{meth_mb}, time-dependent density function theory (TDDFT) \cite{meth_dft}, and semiconductor Bloch equations \cite{meth_sbe}. Impressive results in excellent agreement with experiments were achieved by these methods, see e.g. \cite{dft_show}. Unfortunately, these methods impose very significant numerical effort, determined by the required accuracy of the Green function for many-body perturbation equation, resolution in time and space for the TDDFT and resolution in the Bloch vector space and/or number of energy bands for the semiconductor Bloch equations. In addition, the above exact or semi-exact methods suffer from the necessity to provide parameters which are not readily available a priori, such as pseudopotentials for the TDDFT and transition dipole momenta for the semiconductor Bloch equations.

On the other hand, there exists a class of simple analytical expressions for photoionization rates, which depend only on a few solid state parameters (most typically the band gap) and the instantaneous electric field. They include a family of methods such as Amosov-Delone-Krainov (ADK) \cite{meth_adk} and Perelomov-Popov-Terent'ev PPT \cite{meth_ppt} formalisms, as well as so-called Ivanov-Yudin (IY) \cite{meth_iy} approach. A simple multiphoton expression for the ionization rate proportional to a high power of the electric field \cite{meth_mp} also belongs to this class. These approaches operate on the assumption that the solid state material is equivalent to a dense "gas" of single atoms or molecules, each of which is photoionized independently. Such a strong assumption means that application of simple analytical formulae for the solid-state case is not rigorously grounded, in contrast to their application for the gas phase. 

The analytical methods possess an undoubted advantage of being extremely efficient numerically. They allow to perform high-load tasks such as propagation simulation and optimization without the need of supercomputer systems and/or excessive numerical effort, and therefore have a significant practical importance. On the other hand, the choice of the best analytical formalism, as well as the choice of the normalisation coefficients remains more an art than an exact science. 

\begin{figure}
\includegraphics[width=0.5\textwidth]{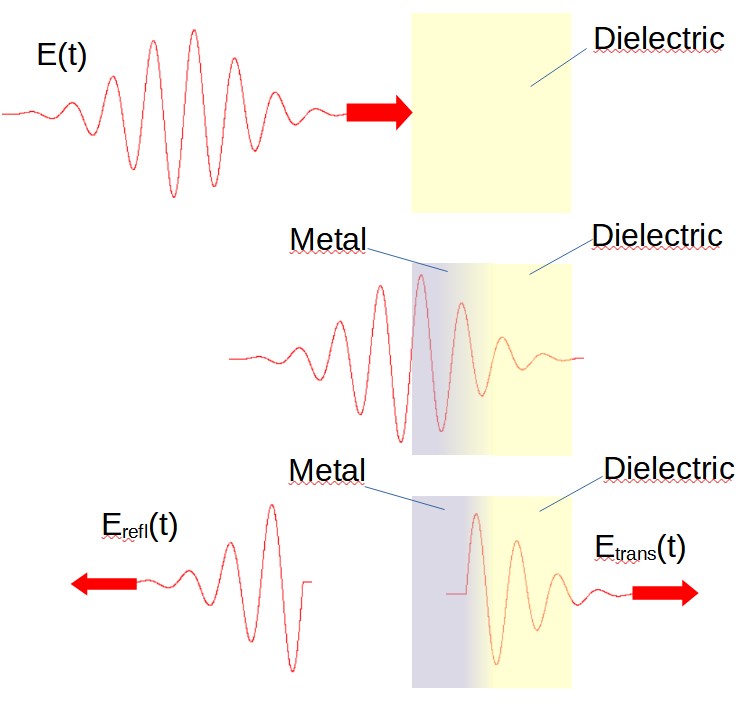}
\caption{Scheme of the proposed setup. In (a), the electric pulse $E(t)$ is normally incident on a dielectric sample, indicated by yellow. In (b), photoionization leads to metallization of the surface layer, indicated by a transition from yellow to blue. In (c), an asymmetric reflected pulse $E_{refl}(t)$ carries information about the photoionization dynamics. }
\end{figure}

We propose an approach for the optimum choice of the photoionization formalism, based on the waveform and spectrum of the pulse which is reflected from a dielectric sample due to photoionization and metallization, as shown by the scheme in Fig. 1. A strong relatively short (sub-100-fs) pulse is normally incident on a dielectric sample [Fig. 1(a)]. After the front of the pulse penetrates the surface, it develops a high concentration of electrons in the conduction band, and its optical response becomes metal-like [represented by the blue color in Fig. 1(b)]. The tail of the incident pulse will be reflected from this metallized layer, carrying a signature of the photoionization process. The spectral amplitudes and phases of the reflected pulse can be calculated using the different analytic expressions (in conjunction with the propagation equations) and compared to the spectral amplitudes and phases of the experimental reflected pulse. Alternatively, instead of the latter, one can compare to the pulse calculated from accurate first-principle methods such as TDDFT; in this case, one can also directly compare the temporal profiles of the reflected pulses.

In this manuscript, we determine the best analytical photoionization model for diamond and visible/near-IR pump, based on the waveform of the reflected pulse benchmarked by a TDDFT simulation. We find that the IY model is the most suitable for the considered set of parameters. For the finite-difference time domain (FDTD) simulations, we have developed a software package PhotoIonizaion-Governed Light Emulation Tool (PIGLET) and made this software freely available to the community. In addition, we show that for very short sub-10-fs pulses a simplistic semi-classical model fails to provide adequate description for any analytical photoionization model. 

The manuscript is structured as follows: after the introduction, in Section 2 we present the theoretical and numerical approach and the PIGLET software. In Section 3, we perform the benchmarking of the analytical formulae for the range of intensities and wavelengths and find the most suitable analytic photoionization model. In Section 4 we consider the case of short sub-10-fs pump pulses, followed by the conclusions.

\section{Theoretical and numerical model}

We consider the normal incidence of linearly polarized pulse, which travel in the direction $+z$, on a semi-infinite dielectric sample. We write the one-dimensional Maxwell equations which govern the time evolution of both the electric $E_x(z,t)$ and the magnetic $H_y(z,t)$ fields during the pulse propagation as follows:
\begin{eqnarray} 
&&\epsilon_0\frac{\partial E_x(z,t)}{\partial t}=-\frac{\partial H_y(z,t)}{\partial z}-J_{ion}(z,t)-\frac{\partial P_x(z,t)}{\partial t}\\
&&\mu_0\frac{\partial H_y(z,t)}{\partial t}=\frac{\partial E_x(z,t)}{\partial z},
\end{eqnarray}
where $P_x(z,t)$ is the polarization and $J_{ion}(z,t)$ is the current which describes the loss due to photoionization \cite{ion}: $J_{ion}(z,t)=N_0[1-\rho(z,t)]\Gamma(z,t)E_g/E_z(z,t)$, whereby $N_0$ is the concentration of the entities, such as atoms or molecules, which can be ionized, $\rho(z,t)$ is the relative density of the ionized entities, $\Gamma(z,t)$ is the ionization rate, and $E_g$ is the bandgap. The material response and motion of the photoionized electrons in the conduction band is described by polarization $P_x(z,t)$, which is connected to the electric field $E_x(z,t)$ in the frequency domain by
\begin{equation}
P_x(z,\omega)=E_x(z,\omega)\left[\epsilon_\infty(\omega)-\frac{N_0\rho(z,t) e^2}{\epsilon_0m_e\omega(\omega-i\nu)}\right]
\end{equation}
where $e$ is the electron charge, $\nu$ is the decay rate of the electron motion, $\epsilon_\infty(\omega)$ is the dielectric susceptibility of the unperturbed material, and $m_e$ is the effective electron mass. Note that the above equation contains $\rho(z,t)$ and therefore is, strictly speaking, incorrect as it mixes time and frequency domains; the exact and correct realization will be given below by Eqs. (4), (10), (11).

The temporal evolution of the relative plasma density is described by 
\begin{equation}
\frac{\partial \rho(z,t)}{\partial t}=[1-\rho(z,t)]\Gamma(z,t).
\end{equation}

For the ionization rate $\Gamma(z,t)$, three formalisms are included. In the Ivanov-Yudin formalism, the cycle-resolved ionization rate is given \cite{meth_iy} by (in atomic units, that is, with frequency $\omega$, time $t$ and field $E$ measured in the corresponding Hartree units $\omega_a=0.26$ rad/as, $t_a=24.2$ as, $x_a=0.0529$ nm, and $E_a=514.2$ V/nm):

  \begin{equation}
    \label{eq:YI}
    \Gamma(z,t)=\frac{\pi}{\tau_{T}}\exp\left(-\sigma_{0}\frac{
      \langle 2E(z,t)^{2}\rangle}{\omega^{3}}\right)\left[\frac{2\kappa^{3}}{
       \sqrt{\langle 2 E(z,t)^{2}\rangle}}\right]^{2Z/\kappa}\exp\left[-\frac{ E(z,t)^{2}}{2\omega^{3}}\sigma_{1}\right].
  \end{equation}

Here $Z$ is the effective atomic charge assumed one for diamond, $\kappa=\sqrt{2E_g}$, $\tau_T=\kappa/\sqrt{\langle E(z,t)\rangle^2}$,
$\sigma_{0}=\frac12(\gamma^{2}+\frac12)\ln
C-\frac12\gamma\sqrt{1+\gamma^{2}}$,
$C=1+2\gamma\sqrt{1+\gamma^{2}}+2\gamma^{2}$, 
$\sigma_{1}=\ln C$-$2\gamma/\sqrt{1+\gamma^{2}}$, and $\gamma=\omega\tau_T$. The quantity $\langle E(z,t)^{2}\rangle$ is the 
averaged value of the squared electric field over few past periods (two periods are assumed here).

In the multiphoton formalism, the $\Gamma(z,t)$  is given (in atomic units) by

  \begin{equation}
    \Gamma(z,t)\sim E(z,t)^{2N_{ph}},
  \end{equation}

where $N_{ph}$ is the lowest number of pump pulse photons which have energy sufficient to overcome the bandgap $E_g$, and a field-independent normalization coefficient is applied.

Finally, in the ADK approximation, the ionization rate is expressed \cite{meth_adk}(in atomic units) as 
  \begin{equation}
    \Gamma(z,t)=\left(\frac{3e'}\pi\right)^{3/2}\frac1{3n^3(2n-1)}\left(\frac{4e'}{(2n-1)|E(z,t)|}\right)^{2n-1.5}\exp\left(-\frac{2}{3n|E(z,t)|}\right),
  \end{equation}
  where $e'=2.71828\dots$ and $n=1/\sqrt{2E_g}$.

  Each of the above models is corrected by a central-wavelength-dependent prefactor, which is selected to match the reflection coefficient to that of the TDDFT calculation. 
  This prefactor is defined in terms of the damage threshold, i.e., the fluence value for which plasma frequency $\omega_p=\sqrt{N_0e^2\rho(0,+\infty)/(\epsilon_0m_e)}$ is equal to the pump frequency \cite{app_our}.

\begin{figure}
\includegraphics[width=0.5\textwidth]{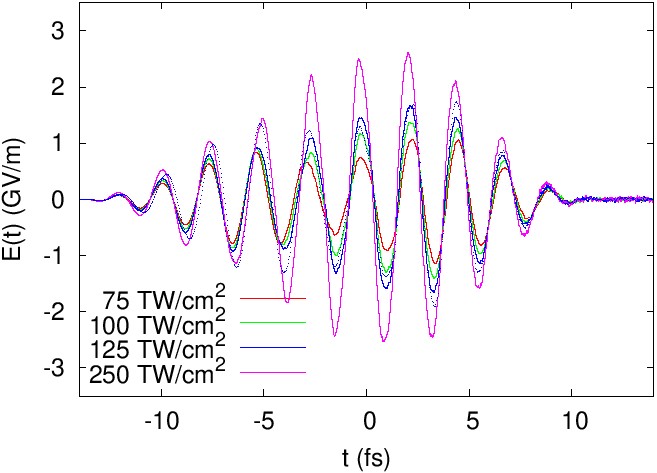}
\caption{Intensity dependence of the reflected pulse as simulated by TDDFT. We consider $\cos^2$-shaped pulses with a full-width half-maximum (FWHM) duration of 12.25 fs and central wavelength of 700 nm, with intensities as indicated in the figure, normally incident on a 40-nm-thick diamond sample. Thin blue dotted line indicates the reflected pulse for pump intensity of 125 TW/cm$^2$ and 80-nm-thick sample.}
\end{figure}

The pulse propagation is described using the standard Yee-cell one-dimensional FDTD approach \cite{FDTD} with he initial-condition approach.
The absorbing boundary conditions at boundaries $z=0$ and $z=z_{max}$ (the boundary of the computation domain in space) are given by 

\begin{eqnarray}
&&c\frac{\partial E_x(0,t)}{\partial z}=\frac{\partial E_x(0,t)}{\partial t}\\
&&c\frac{\partial E_x(z_{max},t)}{\partial z}=-\sqrt{\epsilon(z_{max},t)}\frac{\partial E_x(z_{max},t)}{\partial t},
\end{eqnarray}

whereby electron plasma contribution was taken into account in the time-dependent dielectric function $\epsilon(z_{max},t)$ in Eq. (9).

Finally, the implementation of Eq. (3) in the time domain is based on the following discretization scheme \cite{polarization} which provides the values of $E_x(z,t+dt)$ and $P_x(z,t+dt)$:

\begin{eqnarray}
&&P_x(z,t+dt)-P_x(z,t)+\frac12\nu dt(P_x(z,t+dt)+P_x(z,t))=\nonumber\\&&-0.5\sigma(z,t)\epsilon_0dt(E_x(z,t)+E_x(z,t+dt)),\\
&&\epsilon_\infty(E_x(z,t+dt)-E_x(z,t))+P_x(z,t+dt)-P_x(z,t)=\nonumber\\ &&\frac{dt}{\epsilon_0}(-H_y(z+dz,t)+H_y(z,t)-0.5\sigma(z,t)(E_x(z,t)+E_x(z,t+dt)),
\end{eqnarray}

where $dt$ is the time step and $\sigma=\epsilon_0\omega_p(z,t)^2/\nu=N_0e^2\rho(z,t)/(\nu m_e)$.

The PIGLET code, together with the corresponding detailed documentation, source code, manual and so on, is made available for the community free of charge \cite{github}.

For the aims of benchmarking, we have used the TDDFT SALMON package \cite{salmon_gen}, which is capable of accurate simulation of photoionization-governed propagation as described by FDTD in multiscale approach \cite{salmon_multisc}. SALMON software directly provides the pulse reflected from a dielectric sample, which we consider to be accurate and close to the experimental reflected pulse and suitable for benchmarking, in view of high accuracy of the SALMON code. 12x8x12 points in the Bloch vector space and 16x16x16 points in the real space were used, with a time step of 0.4 attoseconds. The sufficient convergence of the TDDFT simulation was directly checked by repeating the simulations with a higher number of points in space and $k$-space on Fugaku supercomputer, one of the the most powerful supercomputers in the world. 

For the numerical simulations, in the case of diamond, the following values of the above parameters were used: $Z=1$, $E_g=5.5$ eV\cite{par_eg}, $m_e=0.43m_0$\cite{par_me} where
$m_0$ is the mass of a free electron, $N_0=1.75\times10^{29}$ m$^{-3}$ as calculated from the diamond density and the assumption that each carbon atom can be ionized, $\epsilon_\infty(\omega)=5.85$\cite{par_eps}, and $\nu=0.1$ fs$^{-1}$ was assumed.

\section{Numerical results}

Let us start with the temporal profiles of the reflected pulses as calculated by TDDFT, presented in Fig. 2. We consider 12.25-fs FWHM input pulses  with central wavelength of 700 nm and varying intensity incident on a 40-nm-thick diamond sample. For a lower intensity of 75 TW/cm$^2$ (red curve), metallization is not reached, with final relative electron density below the critical relative electron density of $\rho_{crit}=\epsilon_{\infty}\omega^2m_e\epsilon_0/(N_0e^2)$ of 0.0309. Therefore the reflected pulse is mostly symmetric, with its intensity determined by the refractive-index contrast between the vacuum and diamond, corresponding to the energy reflection coefficient of about 0.17. For a higher pump intensity of 100 TW/cm$^2$ and in particular 125 TW/cm$^2$, one can see a pronounces asymmetry of the reflected pulse, with the tail being significantly stronger. This is explained by the metallization (dielectric function becoming negative) close to the center of the pulse. The reflection coefficient rises to around 0.3 for these cases. Finally, for a very high intensity of 250 TW/cm$^2$, the metallization $\rho=\rho_{crit}$ is reached almost immediately, and most of the pulse is reflected, with a reflection coefficient around 0.45. It should be noted that such values of the reflection coefficient in fact mean that much more energy was reflected than transmitted. The reason is that the remaining part of the energy is mostly not transmitted but rather used up to move the electrons across the band gap, and is thus absorbed. The thin dotted line in Fig. 2 indicates the reflected pulse for pump intensity of 125 TW/cm$^2$ but with double sample thickness of 80 nm. One can see that its deviation from the case of 40-nm-thick sample is small, except for a narrow temporal range around -5 fs. We conclude that, for the considered intensity and plasma concentration, only a thin front layer of the sample is mostly responsible for the reflection. 

Based on these findings, we have chosen the case of 125 TW/cm$^2$ with the most clearly pronounced asymmetry for further investigation.

\begin{figure}
\includegraphics[width=0.5\textwidth]{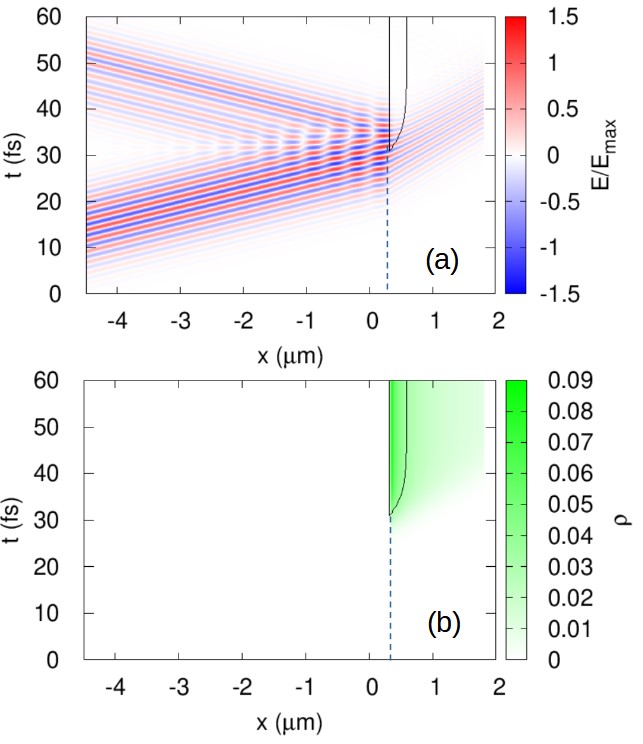}
\caption{Simulation of the pulse reflection based on IY photoionization model and FDTD approach. In (a), the spatiotemporal map of the electric field, and in (b), a map of the relative plasma density, are presented. We consider $\cos^2$-shaped pulse with a FWHM duration of 12.25 fs, central wavelength of 700 nm, and intensity of 125 TW/cm$^2$. The sample surface is at $x=300 nm$, and the metallized reflecting spatiotemporal domain is indicated by a black thin line.}
\end{figure}

\begin{figure}
\includegraphics[width=0.5\textwidth]{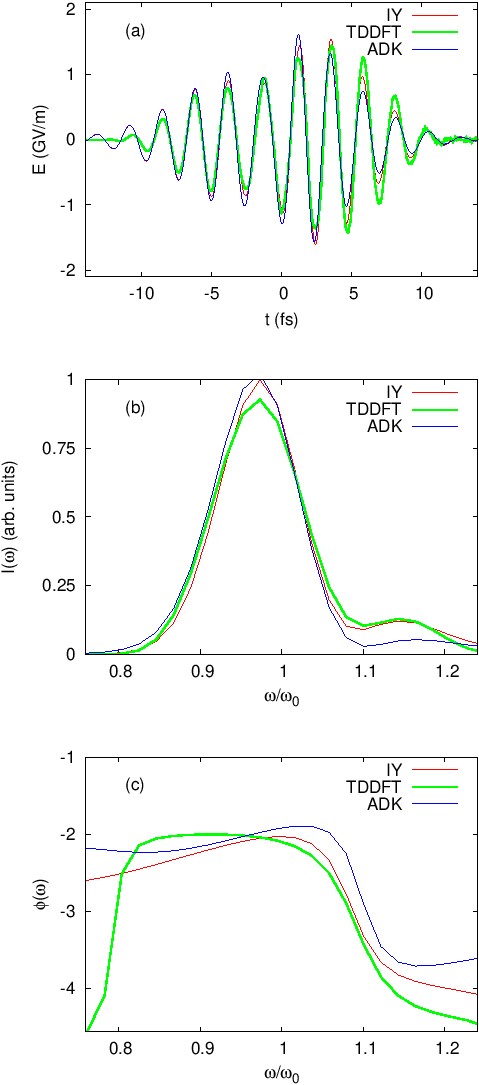}
\caption{Comparison of the results obtained by TDDFT and different analytical models, as indicated in the figure. The temporal profiles (a), spectra (b), and phases (c) are shown. We consider $\cos^2$-shaped pulse with a FWHM duration of 12.25 fs, central wavelength of 700 nm, and intensity of 125 TW/cm$^2$. In (b) and (c), $\omega_0$ is the central pump frequency.}
\end{figure}

For the FDTD method and analytical IY simulation model, the results are shown in Fig. 3. In Fig. 3(a), one can see that the incident pulse (going from left to right) hits the surface of dielectric (indicated by a dashed line) at $x=$0.3 $\mu$m. The front of the pulse penetrates the dielectric, inducing photoionization and leading to metallization of the material. The tail of the pulse is reflected, and thus the reflected pulse has asymmetric form with suppressed front. In Fig. 3(b), the spatiotemporal shape of the induced relative plasma density is shown. One can see the fast formation of plasma close to the center of the pulse around $t=$ 32 fs, with metallized domain (negative dielectric function) shown by a solid black curve.

Let us compare the performance of the two above mentioned simple photoionization models (ADK and IY) with regards to predicting the characteristics of the reflected pulse. The results of the multiphoton model are similar to those of ADK and are not shown here for the sake of clarity. No post-simulation adjustment of the simulated reflection pulse was made. In Fig. 4(a) we show the reflected pulse as simulated by TDDFT (red curve) as well as ADK (blue curve) and IY (red curve) models. One can see that while all models reproduce asymmetry of the reflected pulse, ADK model provides the best agreement. In order to access this agreement quantitatively, we define a figure of merit 
\begin{equation}
\Delta F=\frac{\int_{-\infty}^{\infty}[E_{TDDFT}(t)-E_{model}(t)]^2dt}{\int_{-\infty}^{\infty}[E_{TDDFT}^2(t)+E_{model}^2](t)dt}
\end{equation}
which takes a zero value if the pulses coincide and a value of 2 if they differ by sign. For the case of Fig. 4 we get a FOM of 0.044 for the IY model, 0.085 for the multiphoton model, and 0.091 for the ADK model, clearly indicating a better performance of the IY model by a factor of roughly 2. The better performance of the IY model is related to the fact that it can reproduce both low-intensity (multiphoton) and high-intensity (tunneling-like) regimes of photoionization.

From the experimental perspective, accessing the pulse shape is more involved than characterising the spectral shape and phases. Therefore in Fig. 4(b),(c) we compare the spectral profiles and phases of the TDDFT-simulated reflected pulse and of the two pulses calculated by two analytic models. As expected, one can see that IY provides the best fit also
for the spectral shape and spectral phases, except for the phases in the spectral range below 0.87 of the pump frequency. 

\begin{figure}
\includegraphics[width=0.5\textwidth]{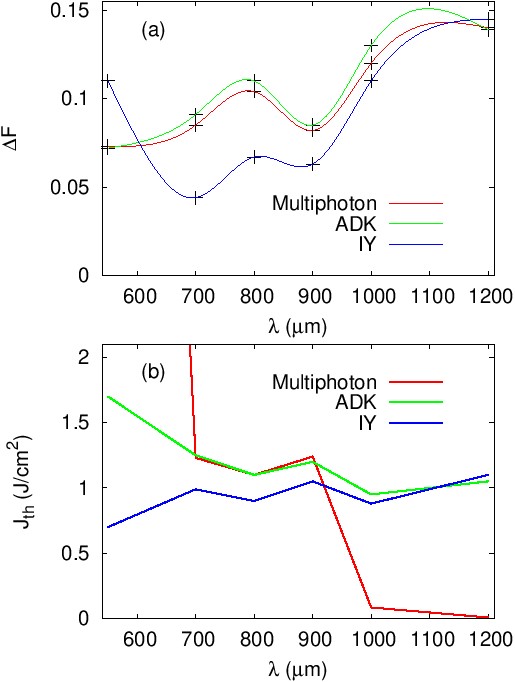}
\caption{Wavelength dependence of the figure of merit $\Delta F$ and the damage threshold $J_{th}$ for the different analytical photoionization models. We consider $\cos^2$-shaped pulse with a FWHM duration of 12.25 fs. Intensity is chosen for each wavelength to provide a energy reflection coefficient of 0.3; this value also provides the most clearly pronounced asymmetry in considered cases.}
\end{figure}

The above finding about the superiority of the IY model in the case of diamond would be of little use and predictive power if it were valid only for a single point in the parameter space. Therefore in Fig. 5 we show the dependence  on the pump wavelength of the $\Delta F$ for the three analytical models. It is clear that the advantage of the IY model is consistent over the whole wavelength range, except for below 600 nm.  Between the multiphoton and ADK models, the figures of merit are similar, however, the multiphoton model requires a strongly wavelength-dependent damage threshold as shown by red curve in Fig. 5(b), which indicates that ADK model is slightly better than multiphoton model, but still clearly worse than the IY model. Also we emphasise that the used value of the damage threshold is in good agreement to the experimental value of 1-3 J/cm$^2$\cite{dia_th}. The above conclusion is established for diamond, however, we speculate that it might be valid also for other wide-gap dielectrics.

\section{Short pulse excitation}

\begin{figure}
\includegraphics[width=0.5\textwidth]{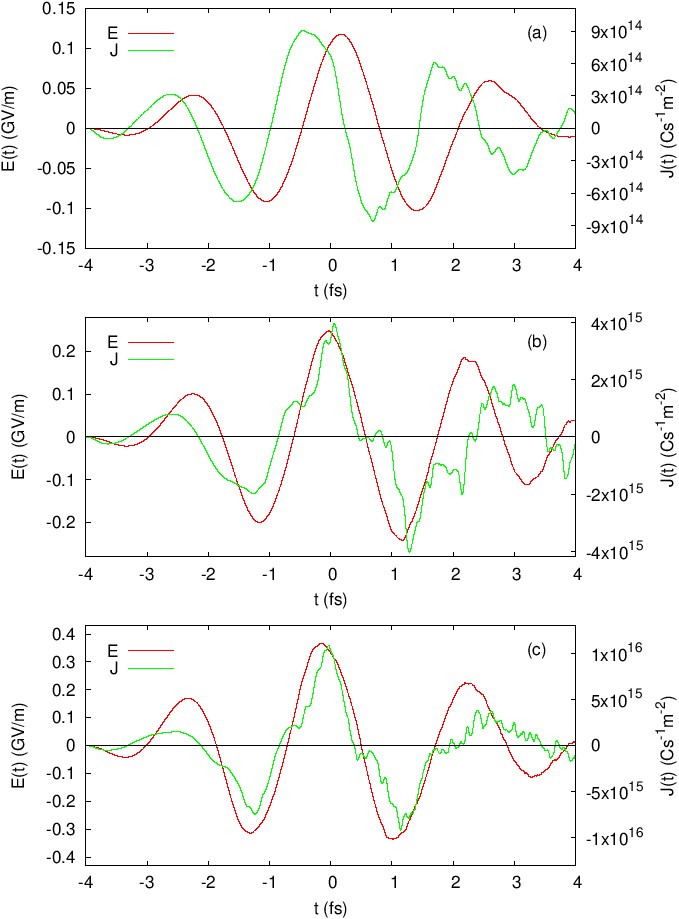}
\caption{Electric field (red curves) and current density (green curves) at the sample surface for different  intensities, as calculated by TDDFT. We consider a $\cos^2$-shaped pulse with a FWHM duration of 4 fs, a central wavelength of 800 nm, and an intensity of 40 TW/cm$^2$ (a), 250 TW/cm$^2$ (b), and 1000 TW/cm$^2$ (c).}
\end{figure}

The interaction of short, sub-10-fs pulses with dielectrics is of particular interest due to potential for applications such as material modification, and therefore deserves a separate study in the framework of this paper. In order to track the photoionization and metallization dynamics, we will analyze the temporal profiles of the electric field $E(t)$ and the induced current density $J(t)$. For dielectrics, $P(t)=\epsilon\epsilon_0E(t)$, and the current density $J=dP/dt\sim dE(t)/dt$ has a phase delay of $-\pi/2$ relative to the field. When the real part of the dielectric function is close to zero, the dynamics is determined by its imaginary part, and the material act as a conductor with a finite real-valued conductivity: $J(t)\sim E(t)$, with current density and electric field in phase. Finally, metals are characterized by a dielectric function with large and negative real part, and  dynamics of the conduction-zone electrons close to being governed by the second Newtons's law: $dJ/dt\sim E$, so that current density has a phase delay of $+\pi/2$ relative to the electric field. In Fig. 6, we analyze the phase delays of the current density at the sample surface, as calculated by TDDFT, for different pump intensities. For the low-intensity case shown in Fig. 6(a), the photoionization is weak, and current density has a delay of roughly $-\pi/2$ relative to the field, corresponding to dielectric. For a higher intensity of 250 TW/cm$^2$, as illustrated in Fig. 6(b), the phase difference shifts from $-\pi/2$ at the beginning of the pulse to roughly zero in the center of the pulse, corresponding to real-conductivity conductor, although for this case one would expect some metallization at the center of the pulse. Also, at this intensity, high-frequency oscillations of the current appear, corresponding to harmonic generation. Quite surprisingly, even at a very high pulse intensity of 1000 TW/cm$^2$, as exhibited in Fig. 6(c), the sample metallizes almost immediately, but we do not observe transition of the relative phase to $+\pi/2$. This is in a seeming contradiction with the fact that after the pulse the relative plasma density is well above the critical value of metallization of 0.025. We explain the persistent real-conductivity behavior of diamond for short-pulse excitation by the interband contributions to the current density as follows: We schematically write the electron wavefunction as $|\psi\rangle=a(t)|v\rangle+b(t)|c\rangle$, where $|v\rangle$ and $|c\rangle$ are the wavefunctions in valence and conduction bands, correspondingly, and note that $a(t)\simeq 1$ and $b(t)\sim t$ for short times. The interband contribution is proportional to $a(t)b(t)\sim t$, while the intraband contribution proportional to $|b(t)|^2\sim t^2$ and is relatively weaker for short pulses. Intraband contribution corresponds to the semi-classical paradigm of almost-free conduction electrons which was considered in the previous section of the manuscript. The results shown in Fig. 6 indicate that this approach alone is no longer sufficient for very short pulses, and interband contribution dominates the current. 

\begin{figure}
\includegraphics[width=0.5\textwidth]{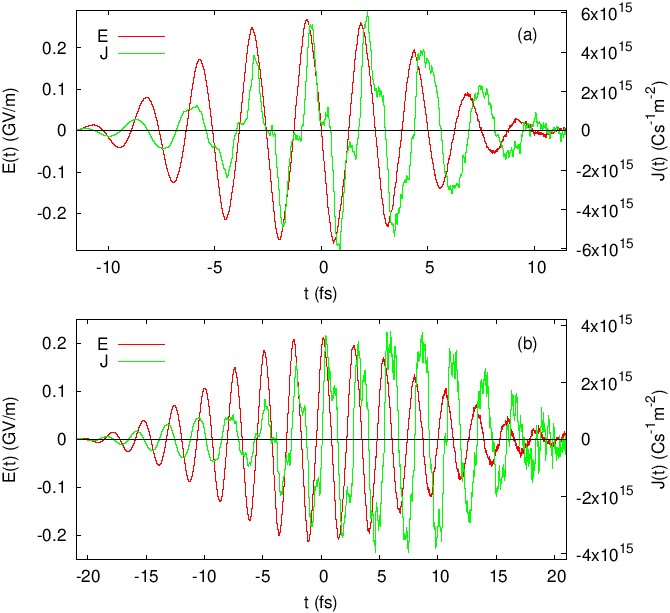}
\caption{Electric field (red curves) and current density (green curves) at the sample surface for different durations, as calculated by TDDFT. We consider a $\cos^2$-shaped pulse with a FWHM duration of 12 fs (a) and 20 fs (b), an intensity of 100 TW/cm$^2$, and a central wavelength of 800 nm.}
\end{figure}

For longer pulses with 12 fs FWGHM, as shown in Fig. 7(a), we have a clear transition from dielectric (for times below -5 fs, phase difference of $-\pi/2$) through real-conductivity behavior (from -5 to 1 fs, phase difference of roughly 0) to metal-like behavior (after time of 1 fs, phase difference is close to $+\pi/2$). In this case, significant population of the conduction zone $b(t)$ develops, and the intraband contribution $b(t)^2$ dominates the interband one. For even longer pulses with FWHM of 20 fs, as shown in Fig. 7(b), we predict a similar transition from dielectric behavior through real-value conductivity to metal-like behavior, in addition, we predict that the maximum of the current is delayed with respect to the pulse. We speculate that this delay might be caused by an onset of non-instantaneous mechanisms of electron transition to the conduction band.

\section{Conclusion}
In this manuscript we have proposed an approach to determine the most suitable photoionization model based on the pulse reflected by the sample due to almost-free electrons in the conduction zone. We compare three analytical models and benchmark them against first-principle TDDFT simulation, analysing the difference between the temporal profiles, spectral amplitudes and spectral phases. We find that for the case of diamond, Ivanov-Yudin model provides the best results. We have developed a software package PIGLET for the FDTD simulation of photoionization-governed propagation; this software is now freely available to the community \cite{github}. In addition, we show that for short sub-10-fs pulses a usual semi-classical model fails to provide adequate description for any analytical photoionization model, and that interband contribution must be included in the analysis.

\begin{acknowledgments}
A.H, Z.R., and K.K. acknowledge financial support from European Union project H2020-MSCA-RISE-2018-823897 "Atlantic".
\end{acknowledgments}

\end{document}